\documentclass{PoS}

\title{A lattice study of light scalar tetraquarks with isopins 0, 1/2 and 1 }

\ShortTitle{A lattice study of light scalar tetraquarks with isopins 0, 1/2 and 1 }

\author{\speaker{Sasa Prelovsek}\\%
        Jozef Stefan Institute and Physics Department at University of Ljubljana, Slovenia\\
        E-mail: \email{sasa.prelovsek@ijs.si}}

\abstract{The observed mass pattern of scalar resonances below $1$ GeV gives preference to the tetraquark assignment over the conventional $\bar qq$ assignment for these states. We present a search for tetraquarks with isospins $0,1/2,1$ in lattice QCD, where isospin channels $1/2$ and $1$ have not been studied before. Our simulation uses Chirally Improved fermions on quenched gauge configurations. We determine three energy levels for each isospin using the variational method. The ground state is consistent with the scattering state, while the two excited states have energy above $2$ GeV. Therefore we find no indication for light tetraquarks at our range of pion masses $344-576$ MeV. }

\FullConference{The XXVI International Symposium on Lattice Field Theory \\
		 July 14 - 19, 2008\\
		 Williamsburg, Virginia, USA}

\begin{document}

\section{Introduction}

\begin{figure}[hb]
\begin{center}
\includegraphics[width=.3\textwidth]{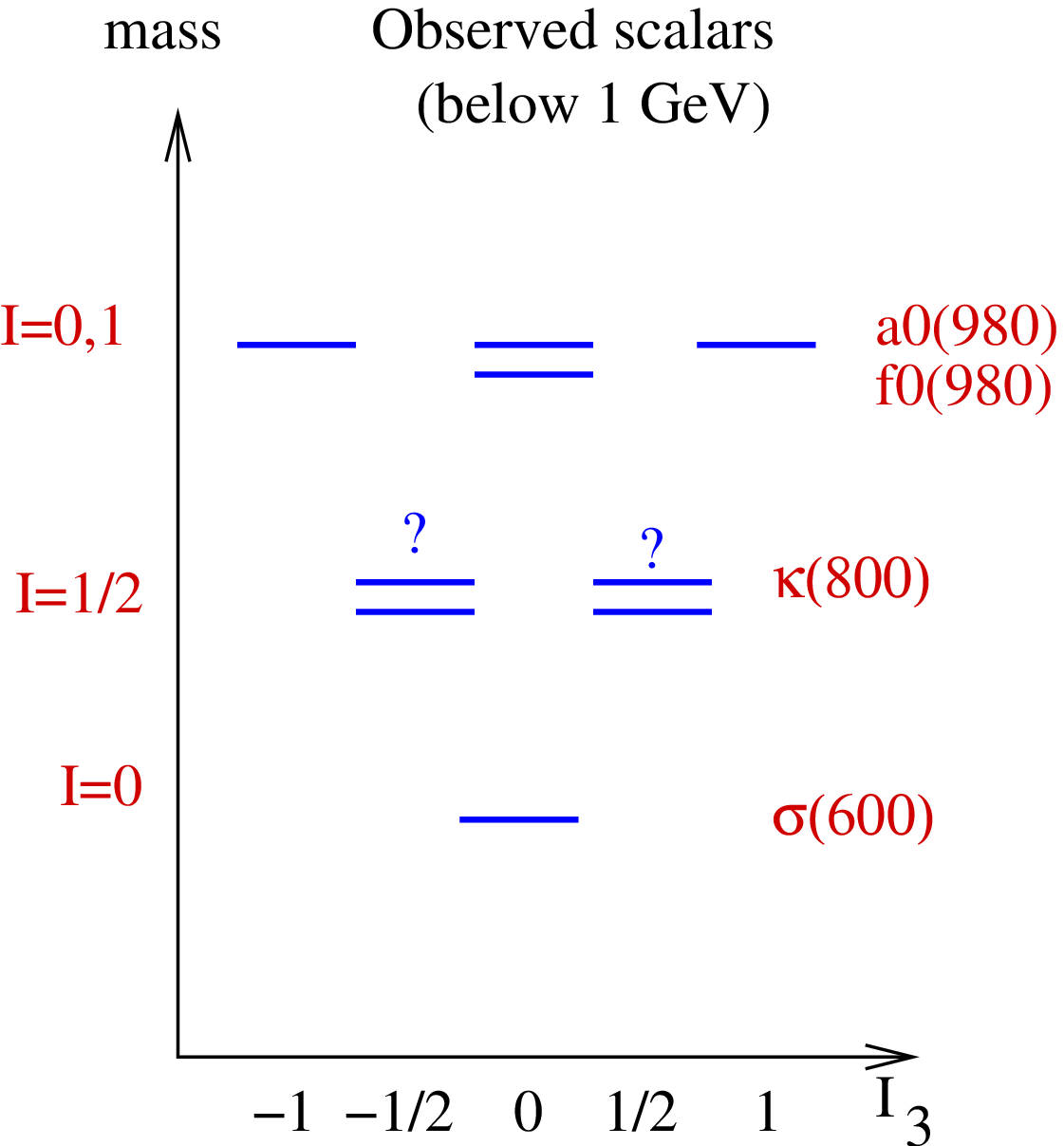}$\quad$
\includegraphics[width=.3\textwidth]{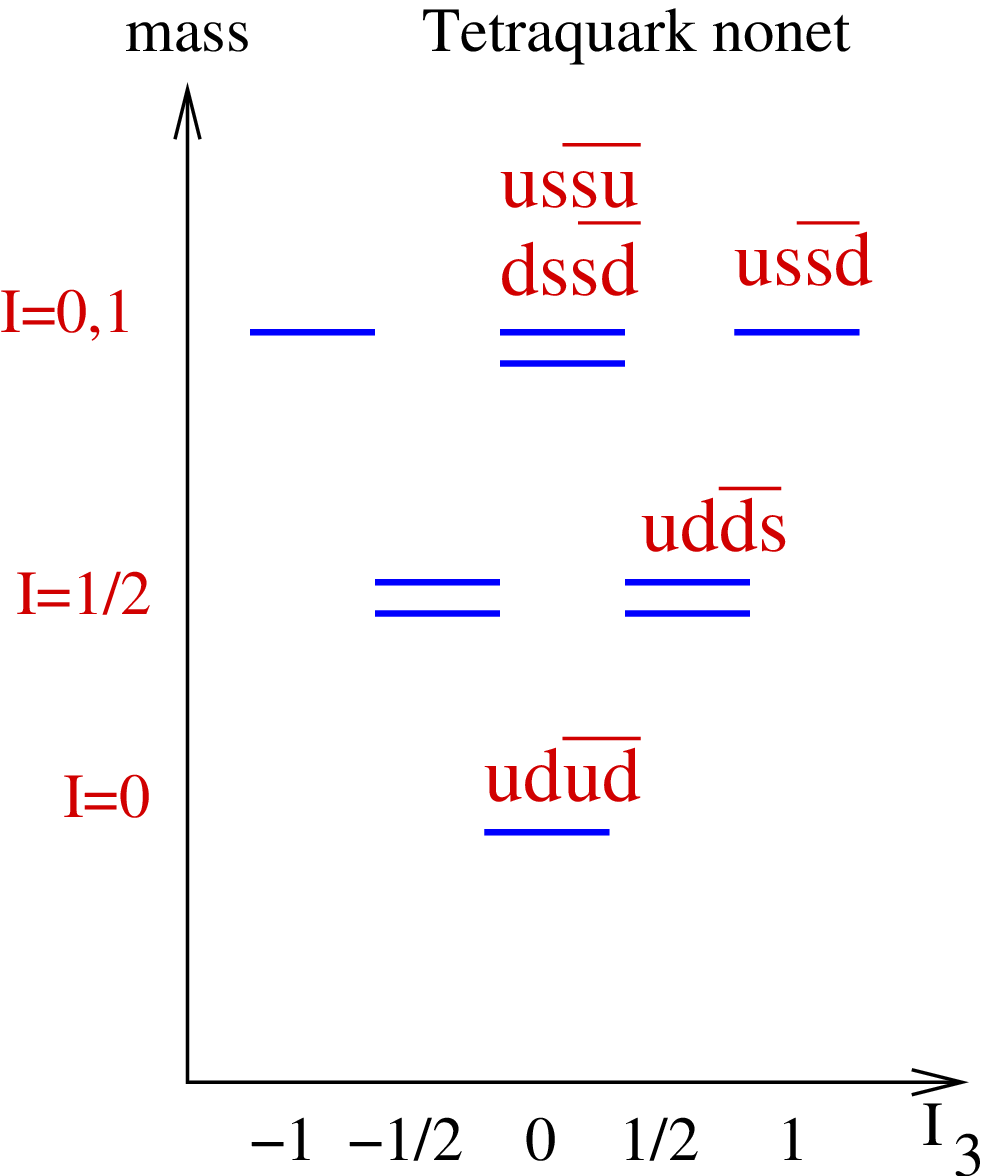}$\quad$
\includegraphics[width=.3\textwidth]{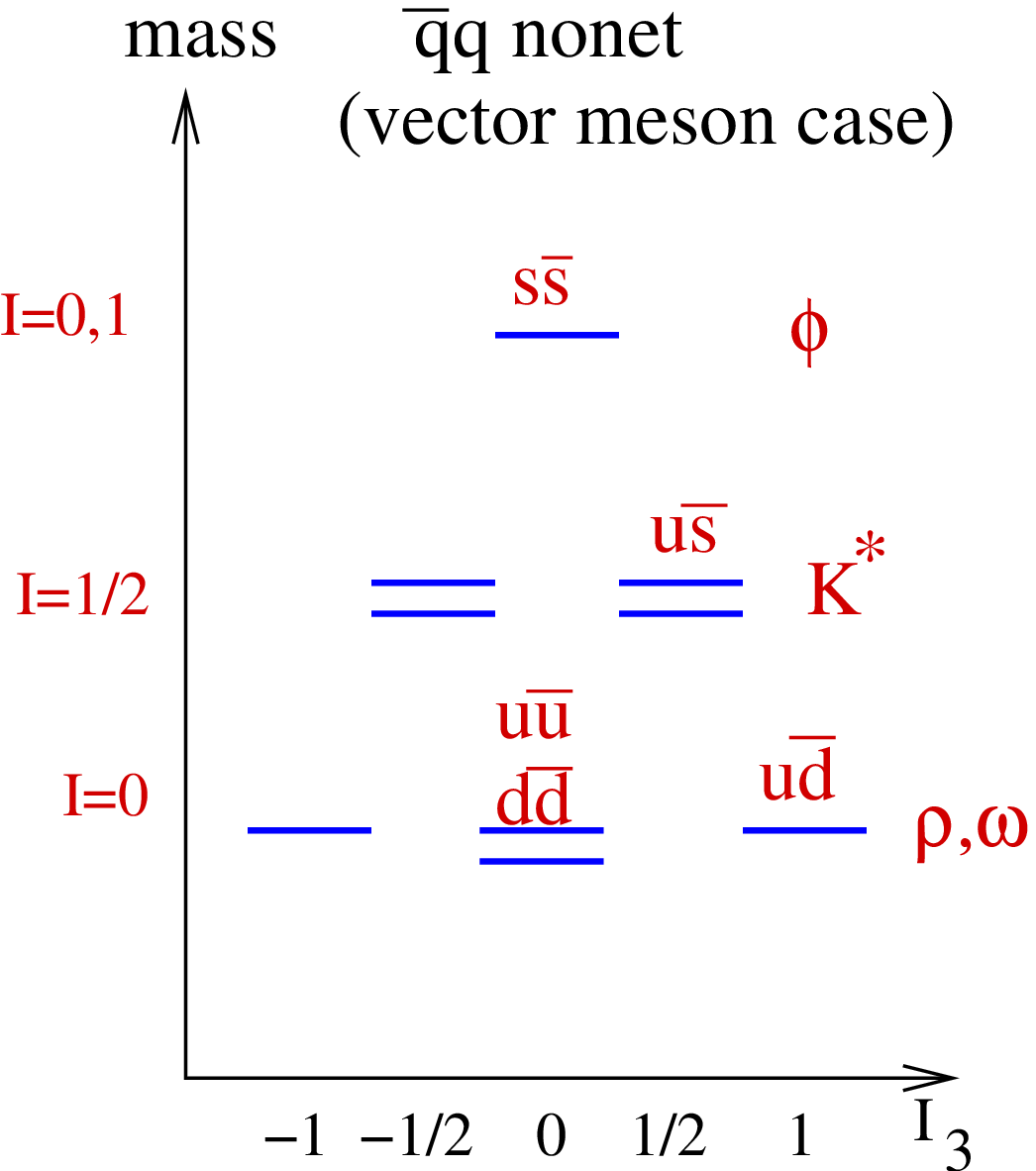}
\end{center}
\caption{ \small Observed spectrum of scalar mesons below $1$ GeV (left), together with the expected spectrum for the nonet of scalar tetraquarks (middle) and  compared with a typical $\bar qq$ spectrum (right).  }\label{fig_pattern}
\end{figure}

The observed mass  pattern of scalar mesons below $1$ GeV, illustrated in Fig. \ref{fig_pattern},  does not agree with the expectations for the conventional $\bar qq$ nonet. The observed ordering $m_\kappa<m_{a0(980)}$ can not be reconciled with the conventional $\bar us$ and $\bar ud$ states since  $m_{\bar us}>m_{\bar ud}$ is expected due to $m_s>m_d$. This is  the key observation which points to the tetraquark interpretation, where light scalar tetraquark resonances   may be formed by combining a ``good'' scalar diquark 
 \begin{equation}
[qQ]_a\equiv \epsilon _{abc} [q_b^T C\gamma_5 Q_c-Q_b^TC\gamma_5 q_c]\quad \mathrm{(color\ and\ flavor\ anti-triplet)}
\end{equation}
with a ``good'' scalar anti-diquark $[\bar q\bar Q]_a$  \cite{tetra_phe}. The 
states $[qq]_{\bar 3_f,\bar 3_c}~[\bar q\bar q]_{3_f,3_c}$ form a flavor nonet of color-singlet scalar states, which are expected to be light. In this case, the $I=1$ state $[us][\bar d\bar s]$ with additional valence pair $\bar ss$ is naturally heavier than the $I=1/2$ state $[ud][\bar d\bar s]$   and the resemblance with the observed spectrum speaks for itself.  

Light scalar tetraquarks have been extensively studied in phenomenological models \cite{tetra_phe}, but there have been only few lattice simulations \cite{jaffe,liu,suganuma}. The main obstacle for identifying  possible tetraquarks  on the lattice is the presence of the scattering  contributions in the correlators.  
All previous simulations considered only $I=0$  and a single correlator, which makes it difficult to disentangle tetraquarks from the scattering. The strongest claim for $\sigma$ as tetraquark was obtained for $m_\pi\simeq 180-300$ MeV by analyzing a single correlator using the sequential empirical Bayes method \cite{liu}. This result needs confirmation using a different method (for example  the variational method used here) before one can claim the existence of light tetraquarks on the lattice with confidence.  

We study the whole flavor pattern with $I=0,1/2,1$ and our goal is to find out whether there are any tetraquark states on the lattice, which could be identified with observed resonances $\sigma(600)$, $\kappa(800)$ and $a_0(980)$.  Our methodology and results are explained in more detail in \cite{tetra_sasa}.

\section{Lattice simulation}

In our simulation, tetraquarks are created and annihilated by diquark anti-diquark interpolators 
\begin{equation}
\label{O_flavor}
{\cal O}^{I=0}=[ud][\bar u\bar d]\ ,\quad 
{\cal O}^{I=1/2}=[ud][\bar d\bar s]\ ,\quad 
{\cal O}^{I=1}=[us][\bar d\bar s]~. 
\end{equation}
In each flavor channel we use three different shapes of interpolators at the source and the sink 
\begin{equation}
\label{O_smear}
{\cal O}_1^I=[q_nQ_n][\bar q_n^\prime \bar Q_n^\prime]\ ,\quad 
{\cal O}_2^I=[q_wQ_w][\bar q_w^\prime \bar Q_w^\prime]\ ,\quad 
{\cal O}_3^I=[q_nQ_w][\bar q_w^\prime \bar Q_n^\prime]~. 
\end{equation}
Here $q_n$ and $q_w$ denote ``narrow'' and ``wide'' Jacobi-smeared quarks. We use exactly the same two smearings as applied in \cite{bgr}, which have approximately Gaussian shape and a width of a few lattice spacings. 
In order to extract energies  $E_n$ of the tetraquark system, we compute the $3\times 3$ correlation matrix for each isospin 
\begin{equation}
\label{cor}
C_{ij}^I(t)=\sum_{\vec x}e^{i\vec p\vec x} \langle 0| {\cal O}_i^I(\vec x,t){\cal O}^{I\dagger}_j(\vec 0,0)|0\rangle_{\vec p=\vec 0}=\sum_n\langle 0|{\cal O}^I_i|n\rangle \langle n|{\cal O}^{I\dagger}_j|0\rangle ~e^{-E_n t}=\sum_n w_n^{ij} e^{-E_n t} ~.
\end{equation}
 
Like all previous tetraquark simulations, 
we use the quenched approximation and discard the disconnected quark contractions. These two approximations allow a definite quark assignment to the states and discard $[\bar q\bar q][qq] \leftrightarrow \bar qq\leftrightarrow vac$  mixing, so there is even a good excuse to use them in these pioneering studies. We work on two\footnote{Equal smearings are used on both volumes.} volumes $V=L^3\times T=16^3\times 32$ and $12^3\times 24$  at the same lattice spacing  $a=0.148$ fm \cite{bgr}. The quark propagators are computed from the Chirally Improved Dirac operator \cite{ci}  with periodic boundary conditions in space and anti-periodic boundary conditions in time. We use $m_la=m_{u,d}a=0.02,~0.04$ and $0.06$ corresponding to $m_\pi= 344,~475$ and $576$ MeV, respectively. The strange quark mass $m_sa=0.08$ is fixed from $m_\phi$.  The analysis requires the knowledge of the kaon masses, which are  $528,~ 576,~620$ MeV for  $m_la=0.02,~0.04,~0.06$.

The extraction of the energies from the correlation functions using a multi-exponential fit $C_{ij}=\sum_n w_n^{ij} e^{-E_n t}$ is unstable. A powerful method to extract excited state energies is the variational method, so we determine the eigenvalues and eigenvectors from the hermitian $3\times 3$ matrix $C(t)$ \footnote{We use the standard eigenvalue problem in order to study $w(L)$. }
\begin{equation}
\label{var}
C(t)\vec v_n(t)=\lambda_n(t)\vec v_n(t)~.
\end{equation}
The resulting large-time dependence of the eigenvalues $\lambda_n(t)=w_n e^{-E_nt} ~[1+{\cal O}(e^{-\Delta_n t})]$ allows a determination of energies $E_{0,1,2}$ and spectral weights $w_{0,1,2}$. The eigenvectors $\vec v_n(t)$ are orthogonal and represent the components of physical states in terms of   variational basis (\ref{O_smear}). 

\section{Results}

Our interpolators  couple to the tetraquarks, if these exist, but they also unavoidably  couple  to the scattering states $\pi\pi$ ($I=0$), $K\pi$ ($I=1/2$) and $K\bar K$, $\pi\eta_{ss}$ ($I=1$) as well as  to the heavier states with the same quantum numbers. The lowest few energy levels of the scattering states $P_1(\vec k)P_2(-\vec k)$ 
\begin{equation}
\label{energies_scat}
E^{P_1(\vec j)P_2(-\vec j)}\simeq m_{P1}+m_{P2},\ ...\ ,\ \sqrt{m_{P1}^2+\biggl(\frac{2\pi \vec j}{L}\biggr)^2}+\sqrt{m_{P2}^2+\biggl(\frac{2\pi \vec j}{L}\biggr)^2},...
\end{equation}
are well separated for our $L$ and we have to identify them before attributing any energy levels $E\simeq m_{\sigma,\kappa,a_0}$ to the tetraquarks.

\begin{figure}[ht]
\begin{center}

\vspace{0.3cm}

\includegraphics[width=.7\textwidth]{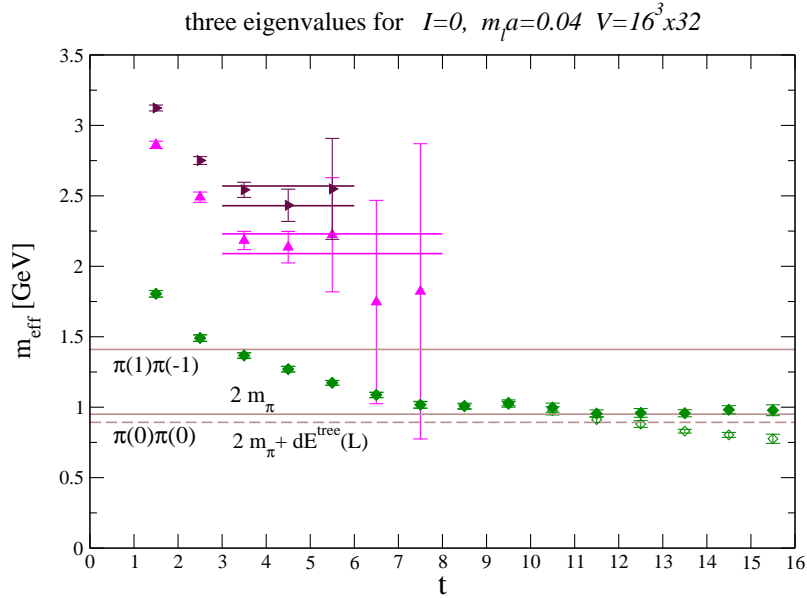} 
\end{center}
\caption{ \small Effective masses for the three eigenvalues $\lambda_{0,1,2}(t)$ at $I=0$, $m_la=0.04$ and $V=16^3\times 32$. They are obtained from conventional cosh-type definition, except for the full diamonds which are obtained from $\lambda_0(t)-const=w_0[e^{-m_{eff}t}+e^{-m_{eff}(T-t)}]$ (see Appendix). The lines give energy levels for  the scattering states: full lines present non-interacting energies, while dashed lines take into account tree-level energy shifts.   }
\label{fig_eigenvalues}
\end{figure}

\begin{figure}[ht]
\begin{center}
\includegraphics[width=.77\textwidth]{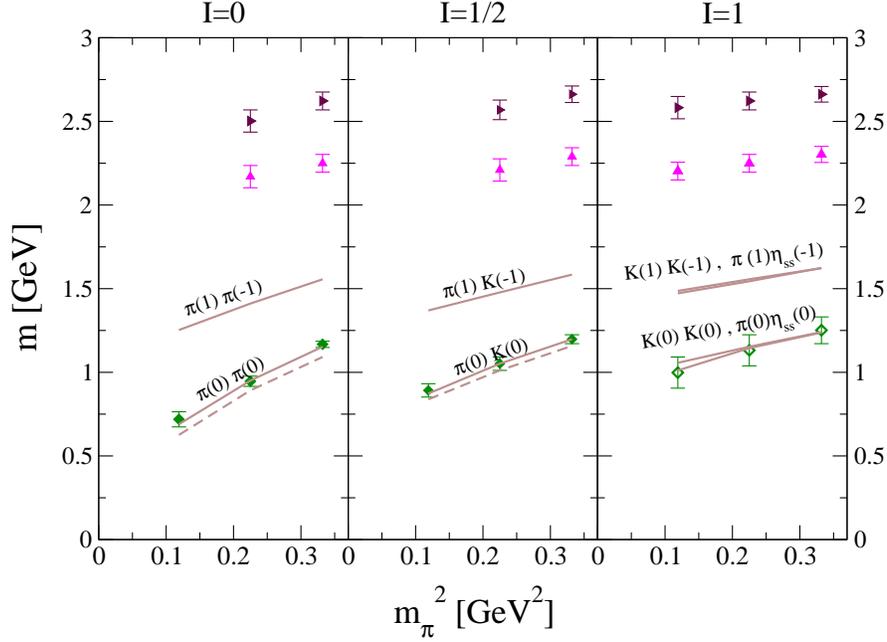}
\end{center}
\caption{ \small  Three lowest energy levels from tetraquark correlators in $I=0,1/2,1$ channels at lattice volume $16^3\times 32$. The lines give analytic energy levels for  scattering states: full lines present non-interacting energies, while dashed lines take into account tree-level energy shifts.    }
\label{fig_spectrum}
\end{figure}

The three energy levels for $I=0$ are represented by the effective masses of the three eigenvalues $\lambda_n(t)$ in Fig. \ref{fig_eigenvalues}. Similar effective masses are obtained for all isospins, quark masses and volumes. The energies  $E_{0,1,2}$ extracted from $\lambda_{0,1,2}(t)$  are summarized in Fig. \ref{fig_spectrum} for all isospin channels. The excited energies were obtained from a conventional two-parameter fit  $\lambda_{1,2}(t)=w_{1,2}[e^{-E_{1,2}t}+e^{-E_{1,2}(T-t)}]$. The fit of $\lambda_0(t)$  (\ref{c_bc}) takes into account a non-standard time dependence at finite $T$ (noticed from decreasing $m_{eff}$ near $t\simeq T/2$) and is described in Appendix.   

\begin{figure}[ht]
\begin{center}

$~$

\vspace{1.2cm}

\includegraphics[width=.55\textwidth]{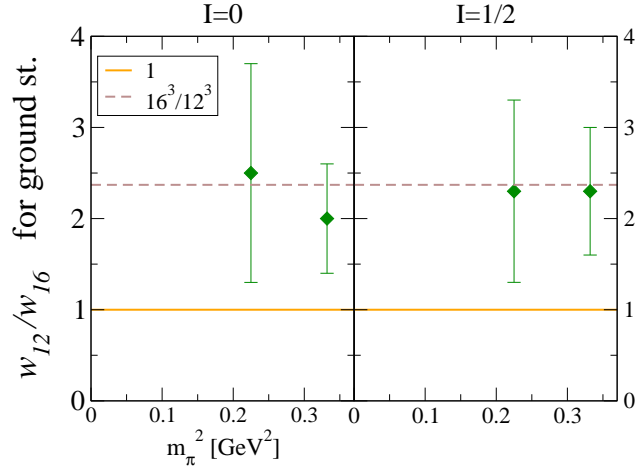}
\end{center}
\caption{ \small The ratio of spectral weights $w_0(L=12)/w_0(L=16)$ 
for $I=0,1/2$ as computed from ground state eigenvalues for two volumes $L^3$.
}
\label{fig_w}
\end{figure}

The {\it ground state} energies in $I=0,1/2$ and $1$ channels are close to $2m_\pi$, $m_\pi+m_K$ and $2m_K,~m_\pi+m_{\eta_{ss}}$, respectively, which indicates that  all ground states correspond to the scattering states $P_1(\vec 0)P_2(\vec 0)$ (see Fig. \ref{fig_spectrum}). Another indication in favor of this interpretation comes from  $w_0(L=12)/w_0(L=16)\simeq 16^3/12^3$ (see Fig. \ref{fig_w}).  This agrees with the expected dependence $w_0\propto 1/L^3$  for scattering states \cite{liu}, which follows from the integral over the loop momenta $\int \frac{d\vec k}{(2\pi)^3} f(\vec k,t)\to \frac{1}{L^3} \sum_{\vec k} f(\vec k,t)$ with $dk_i=2\pi/L$.  The third indication comes from a non-standard time dependence of $\lambda_0(t)$, which agrees with the expected time-dependence of $P_1(\vec 0)P_2(\vec 0)$ at finite $T$ (see Appendix). 

The most important feature of the spectrum in Fig. \ref{fig_spectrum}  is a large gap above the ground state: {\it the first and the second excited states} appear only at  energies above $2$ GeV. Whatever the nature of these two excited states are, they are much too heavy to correspond to $\sigma(600),\ \kappa(800)$ or $a_0(980)$, which are the light tetraquark candidates we are after. The two excited states may correspond to  $P_1(\vec k)P_2(-\vec k)$  with higher $\vec k$ or to some other energetic state. We refrain from identifying the excited states with certain physical objects as such massive states are not a focus of our present study. 

Our conclusion is that we find no evidence for  light tetraquarks at our range of pion masses $344-576$ MeV. This is not in conflict with the simulation of  the Kentucky group \cite{liu}, which finds indication for an $I=0$ tetraquark for pion masses $180-300$ MeV (but not above that) since all our pion masses are just above $300$ MeV. 

\section{The absence of scattering states with $|\vec k|=2\pi/L$ in the spectrum }

Why  are there  no states close to the energies of $P_1(1)P_2(-1)$ with $|\vec k|=2\pi/L$  in the spectrum of Figs. \ref{fig_eigenvalues} and \ref{fig_spectrum}? We  believe  this is due to the fact that all our sources (\ref{O_smear}) have a small spatial extent and behave close to point like. The point source couples to all the scattering states 
equally $\sum_{\vec k} g_{\vec k} |P_1(\vec k)P_2(-\vec k)\rangle$
up to a factor $g_{\vec k}$, which gives  a Lorentz structure of the coupling. 
In this approximation each source ${\cal O}_{1,2,3}$ (\ref{O_smear}) 
couples to the few lowest scattering states  equally (within our error bars) 
\begin{equation}
\label{hypothesis} 
{\cal O}_i=c_i\sum_{\vec k} g_{\vec k}~|P_1 (\vec k)P_2(-\vec k)\rangle +
a_i|a\rangle+b_i|b\rangle\qquad i=1,2,3
\end{equation}
and only the overall strengths $c_{1,2,3}$ are different. We assume also  
that the sources couple  to two other physical states $a$ and $b$. Given these linear combinations, one can construct the corresponding $3\times 3$ correlation matrix and it can be easily shown that its three eigenvalues are 
\begin{equation}
\label{lambda_var}
\lambda_0(t)=w_0 ~\sum_{\vec k} g_{\vec k} ~ \frac{ e^{-(E^{P_1(\vec k)}+E^{P_2(\vec k)})t}}{E^{P_1(\vec k)}E^{P_2(\vec k)} }\ ,\quad 
\lambda_{a,b}(t)=w_{a,b}~e^{-E_{a,b} t}~.
\end{equation} 
The physical states $a$ and $b$ get their own exponentially falling eigenvalues, while a  tower of few lowest scattering states contributes to a single eigenvalue in this approximation. The agreement of  $\lambda_0(t)$ and the 
analytic prediction (\ref{lambda_var}) with $g_{\vec k}=1$ \cite{tetra_sasa} gives us confidence about the hypothesis (\ref{hypothesis}). So our basis does not dissentangle the few lowest scattering states into seperate eigenvalues. They  would contribute to seperate eigenvalues only by using a different or larger basis.

\section{Conclusions and outlook}

Our lattice simulation gives no indication that the observed resonances $\sigma(600)$, $\kappa(800)$ and $a_0(980)$ are tetraquarks. However, one should not give up hopes for finding these interesting objects on the lattice. Indeed, our simulation with pion masses $344-576$ MeV does not exclude the possibility of finding tetraquarks for lighter $m_\pi$ or for  a different interpolator basis. A stimulating lattice indication for $\sigma$ as a  tetraquark  at $m_\pi=182-300$  MeV has already been presented in \cite{liu}. 

\vspace{0.5cm}

{\bf Acknowledgments}

\vspace{0.2cm}

I would like to thank D. Mohler, C. Lang, C. Gattringer, L. Glozman, Keh-Fei Liu, T. Draper, N. Mathur, M. Savage and W. Detmold for useful discussions. This work is supported in part by European RTN network, contract number MRTN-CT-035482 (FLAVIAnet).  

\appendix

\section{Appendix: Effect of finite $T$ on scattering states}

\begin{figure}[h]
\begin{center}
\includegraphics[width=.6\textwidth]{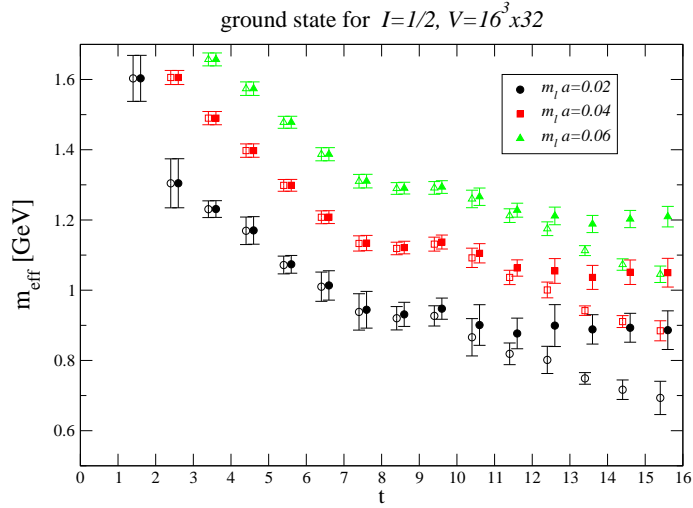}
\end{center}
\caption{ \small The effective masses for the $I=1/2$ ground states at various $m_la$ and $V=16^3\times 32$. Empty symbols are obtained using the conventional cosh-type effective mass $\lambda_0(t)=w_0[e^{-m_{eff}t}+e^{-m_{eff}(T-t)}]$, while the full symbols take into account the correct form $\lambda_0(t)- A [e^{-m_\pi t}e^{-m_K (T-t)} + e^{-m_K t}e^{-m_\pi (T-t)}]=w_0[e^{-m_{eff}t}+e^{-m_{eff}(T-t)}]$ for $K(\vec 0)\pi(\vec 0)$ scattering (see Appendix). }\label{fig_corrected}
\end{figure}

We find that the cosh-type effective mass for the ground state $P_1(\vec 0)P_2(\vec 0)$ is decreasing near $t\simeq T/2$ (empty symbols in Figs. \ref{fig_eigenvalues} and \ref{fig_corrected}), which means 
 that $\lambda_0(t)$ does not behave as $e^{-E_0 t}+e^{-E_0 (T-t)}$ at large $t$.  The time dependence of  $P_1(\vec 0)P_2(\vec 0)$  with  anti-periodic propagators in time is 
\begin{equation}
\label{c_bc}
\lambda_0^{P_1(0)P_2(0)}(t)=w_0 [ e^{-E_0 t}+e^{-E_0 (T-t)}]+ A [e^{-m_{P1} t}e^{-m_{P2} (T-t)} + e^{-m_{P2} t}e^{-m_{P1} (T-t)}]\ ,\quad E_0\simeq m_{P1}+m_{P2}~. 
\end{equation}
In the last term one pseudoscalar propagates forward and the other backward in time  (see Appendix A of \cite{savage}). The ground state energies for isospins $0$ and $1/2$ in Fig. \ref{fig_spectrum} were obtained by fitting  $\lambda_0(t)$ to (\ref{c_bc}) with three unknown parameters $E_0$, $w_0$ and $A$ \footnote{At the time of our presentation at Lattice 2008, we were not aware of the last term in (\ref{c_bc}), so the numerical results on our slides are slightly different. The general physical conclusions concerning the tetraquarks are however the same. }. The effective mass  obtained from $\lambda_0(t)- A [e^{-m_{P1} t}e^{-m_{P2} (T-t)} + e^{-m_{P2} t}e^{-m_{P1} (T-t)}]$ is flat near $t\simeq T/2$ (full symbols in  Figs. \ref{fig_eigenvalues} and \ref{fig_corrected}). 
 
The $I=1$ channel contains $K\bar K$ and $\pi\eta_{ss}$ scattering states  and 
is therefore more challenging. The ground state energies were obtained from 
a naive one-state (cosh) fit since a two-state fit is very unstable. 
 
\end{document}